\documentclass[ijoc,nonblindrev]{informs3} 
\usepackage{color}
\usepackage{url}
\usepackage{hyperref}
 \usepackage{amsmath}
\hypersetup{
    colorlinks=blue,
    linkcolor=blue,
    filecolor=blue,
    urlcolor=blue,
}
\OneAndAHalfSpacedXII 

\usepackage{natbib}
\usepackage[ruled]{algorithm}
\usepackage[noend]{algorithmic}
 \bibpunct[, ]{(}{)}{,}{a}{}{,}%
 \def\bibfont{\small}%
\TheoremsNumberedThrough     

\EquationsNumberedThrough    

\MANUSCRIPTNO{} 

\begin{document}

\TITLE{Optimizing Ad Allocation in Social Advertising}
\ARTICLEAUTHORS{%
\AUTHOR{Shaojie Tang}
\AFF{University of Texas at Dallas, \EMAIL{shaojie.tang@utdallas.edu}}
\AUTHOR{Jing Yuan}
\AFF{University of Texas at Dallas}}

\ABSTRACT{Social advertising (or social promotion) is an effective approach that produces a significant cascade of adoption through influence in the online social networks. The goal of this work is to optimize the ad allocation from the platform's perspective. On the one hand, the platform would like to  maximize revenue earned from each advertiser by exposing their ads to as many people as possible, one the other hand, the platform wants to reduce free-riding to ensure the truthfulness of the advertiser. To access this tradeoff, we adopt the concept of \emph{regret} \citep{viral2015social} to measure the performance of an ad allocation scheme. In particular, we study two social advertising problems: \emph{budgeted social advertising problem} and \emph{unconstrained social advertising problem}. In the first problem, we aim at selecting a set of seeds for each advertiser that minimizes the regret while setting budget constraints on the attention cost; in the second problem, we propose to optimize a linear combination of the regret and attention costs. We prove that both problems are NP-hard, and then develop a constant factor approximation algorithm for each problem.}
\maketitle

\emph{``When we boost posts, we see twice as much engagement, twice as much website traffic and often double our sales."} -- Jaron Schneider, Features Editor, Fstoppers

\section{Introduction}
\label{sec-intro}

Social advertising (or social promotion) is an effective approach that produces a significant cascade of adoption through influence in the online social networks. During a typical social advertising campaign, advertisers attempt to persuade influential consumers to promote their products or services among his friends. With more people using social networking services, recent days have witnessed a boom of social networking sites that offer social advertising (SA) services. To name a few,  the primary SA mechanisms adopted by Facebook are Facebook Ads, Promoted Posts and Boost Posts; Twitter allows businesses to promote their accounts and Tweets as well as promote ``trends"; LinkedIn users can create an advert, sponsor content or use Sponsored InMail to launch an email marketing campaign. Take Facebook as an example, \emph{boosting posts} is considered as  an effective way to get more exposure for one's posts, offers or special events. It allows  businesses to pay for posts to be more predominantly displayed on news feeds. Facebook users will see promoted posts labeled with ``Sponsored" in the news feed (not in the right rail where Facebook ads live) both on desktop and mobile. Promoted posts have the same targeting ability that organic posts do, thus they can propagate across the network through ``reposts" or ``shares". Recent field studies \citep{bakshy2012social}\citep{tucker2012social} find that social advertising is more effective than conventional demographically targeted or untargeted ads.

The goal of this work is to optimize the ad allocation from the platform's perspective. We consider the cost per engagement (CPE) model, the advertiser buy a block of ``engagements" such as impressions or clicks from the platform owner via contracts, and the advertiser pays the platform an amount $\alpha_i$ per engagement that is delivered from its ad $a_i$. Each advertiser also sets his budget $B_i$ that specifies the maximum amount of money he would like to pay. It was worth noting that this budget is fixed regardless of the actual amount of engagements that are received at the end of the campaign. Therefore, due to the uncertainty of virality, it is possible that an advertiser may receive more engagements than would be expected under his budget. Unfortunately, this uncertainty may utterly destroy the truthfulness of the advertiser, i.e., the advertiser tends to declare lower budget, hoping to obtain more engagements. Then it is interesting to observe that, on the one hand, the platform would like to  maximize revenue earned from each advertiser by exposing their ads to as many people as possible, one the other hand, the platform wants to reduce free-riding to ensure the truthfulness of the advertiser. To access this tradeoff, we adopt the concept of \emph{regret} \citep{viral2015social} to measure the performance of an ad allocation scheme.

In addition, as the promoted posts are displayed along with organic posts, it is possible to impede the user experience by pushing too many promoted posts to one user. One way to mitigate this is to set a limit, called \emph{attention constraint}, on the maximum number of promoted posts that can be  pushed to each individual user as well as the whole community.

\textbf{Our Results:} In this paper, we propose and study two social advertising problems: \emph{budgeted social advertising problem} and \emph{unconstrained social advertising problem}. In the first problem, we aim at selecting a set of seeds for each advertiser that minimizes the regret while setting budget constraints on the attention cost; in the second problem, we propose to optimize a linear combination of the regret and attention costs. We first prove that both problems are NP-hard by reducing them from traditional influence maximization problem. Then we develop two constant factor approximation algorithms for each problem.

\section{Preliminaries}
\subsection{Matroid}
A matroid $M = (\Omega, \mathcal{I})$ is defined on a finite ground set $\Omega$,  and $\mathcal{I}$ is a family of subsets of $\Omega$ which are called independent sets.
For $M$ to be a matroid, $\mathcal{I}$ must satisfy two properties:
\begin{itemize}
\item (I1) if $X \subseteq Y$ and $Y \ni \mathcal{I}$ then $X \in \mathcal{I}$,
\item (I2) if $X \in I$ and $Y \in \mathcal{I}$ and $|Y|>|X|$ then $\exists  e \in Y\backslash X: X\cup\{e\}\in \mathcal{I}$.
\end{itemize}
Property (I1) says that every subset of an independent set is independent. Property (I2), which is also called \emph{independent set exchange property}, says that if $X$ and $Y$ are independent sets, and $Y$ has more elements than $X$, there exists an element in $Y\backslash X$ that by adding that element to $X$ gives larger independent set. According to Property (I2), one can verify that all maximal independent sets have the same cardinality. A maximal independent set is called
a base of the matroid.
\subsection{Submodular function}
Consider an arbitrary function $f(S)$ that maps subsets of a finite ground set $\Omega$ to non-negative real numbers. We say that $f$ is submodular if it satisfies a natural ``diminishing returns" property: the marginal gain from adding an element to a set $S$ is at least as high as the marginal gain from adding the same element to a superset of $S$. Formally, a submodular function satisfies the follow property: For every $X, Y \subseteq \Omega$ with $X \subseteq Y$ and every $x \in \Omega \backslash Y$, we have that \[f(X\cup \{x\})-f(X)\geq f(Y\cup \{x\})-f(Y)\]
We say a submodular function $f$ is monotone if $f(X) \leq f(Y)$ whenever $X \subseteq Y$.
\subsection{Propagation Model}
To capture the dynamics of ads propagation in social networks, one of the most widely used model, called
\emph{Independent Cascade Model}, is investigated recently in the
context of marketing \citep{Goldenberg1}\citep{Goldenberg2}\citep{kdd03}. To account for the heterogeneity of ads propagation under different ads, we adopted an extended propagation model, called topic-aware propagation model (TIC). Let $G_i=(\mathcal{V}, p_i(E))$ denote the diffusion graph under ad (or topic) $a_i$, where $\mathcal{V}$ represent the set of all users in the network, $p_i(u,v)$ represents the diffusion probability between $u$ and $v$ for ad $a_i$. TIC describes a spreading process
comprising of seed nodes and non-seed nodes.  The process unfolds in discrete timesteps. In each timestep, when a user $u$ clicks an ad $a_i$, it has one chance of influencing each inactive neighbor $v$ with success probability $p_i(u,v)$.
More formally, the input
to the independent cascade model is an initial set of seed nodes $S_i\subseteq \mathcal{V}$ for each ad $a_i$. Let $\sigma_i(S_i)$ denote the expected number of clicks (or engagements) received from ad $a_i$ under seed set $S_i$. Let $\alpha_i$ denote the cost per-engagement for ad $a_i$, the expected revenue received from $a_i$ is $\alpha_i\cdot \sigma_i(S_i)$. As proved in \citep{kdd03}, $\alpha_i\cdot \sigma_i(S_i)$ is a submodular and monotone function.
\section{Problem Statement}
Given a hyper social graph $\mathcal{G}=(G_1, G_2, \cdots, G_{|\mathcal{A}|})$, where each graph $G_i$ represents the diffusion network under topic or ad $a_i$. 
Assume there are $K$ advertisers  that participate in the campaign, denoted by $\mathcal{A}=\{a_1, a_2, \cdots, a_{|\mathcal{A}|}\}$, each advertiser $a_i$ has a finite budget $B_i$. The host needs to identify and allocate a set of users to each of the advertisers. On the one hand, the platform would like to  maximize revenue earned from each advertiser by exposing their ads to as many people as possible, one the other hand, the platform wants to reduce free-riding to ensure the truthfulness of the advertiser. To access this tradeoff, we introduce the concept of regret to measure the performance of an ads allocation scheme. The regret under seed set $\mathcal{S}=\{S_1, S_2, \cdots, S_{|\mathcal{A}|}\}$ is defined as $\sum_{a_i\in \mathcal{A}}|\alpha_i\cdot \sigma_i(S_i)-B_i|$.
To minimize the regret is equivalent to maximizing the following utility function:
\[U(\mathcal{S})=\sum_{a_i\in \mathcal{A}} U_i(\mathcal{S})\]
where
\begin{equation}
\label{equ:23}
U_i(\mathcal{S})=
\begin{cases}
\alpha_i\cdot \sigma_i(S_i) &\mbox{ if $\alpha_i\cdot\sigma_i(S_i)\leq B_i$}\\
2\cdot B_i - \alpha_i\cdot \sigma_i(S_i)  &\mbox{ if $\alpha_i\cdot\sigma_i(S_i)> B_i$}
\end{cases}
\end{equation}
Any ad allocation can be represented using a $|\mathcal{V}|\times|\mathcal{A}|$ matrix $\mathbf{X}$, called allocation matrix, where
 \[X_{ij}=\begin{cases}
1 & \mbox{if user $v_i$ is assigned to ad $a_j$}\\
0 & \mbox{otherwise}
\end{cases}\]
Then the individual attention cost on user $v_i$ is $\sum_{j=1}^{|\mathcal{A}|} X_{ij}$
and the overall attention cost is $\sum_{i=1}^{|\mathcal{V}|}  \sum_{j=1}^{|\mathcal{A}|} X_{ij}$. In the rest of this paper, we use $\mathcal{S}$ and $\mathbf{X}$ interchangeably to represent an ad allocation.

Selection of good seeds with high utility and small attention cost is a critical decision faced by every platform.  In this paper, we propose and study two problems that allow to combine the two objectives: \emph{Budgeted Social Advertising problem} and \emph{Unconstrained Social Advertising problem}. In the first problem, we aim at selecting a set of seeds for each advertiser that maximizes the utility while setting budget constraints on the attention cost; in the second problem, we propose to maximize a linear combination of the two measures.

\subsection{Budgeted Social Advertising Problem}
In the budgeted social advertising problem, we set hard constraints on individual attention cost and overall attention cost: let $\kappa_i$ denote the individual attention limit of user $v_i$ and $K$ denote the overall attention limit, then $\forall v_i \in \mathcal{V}:\mbox{    } \sum_{j=1}^{|\mathcal{A}|} X_{ij} \leq \kappa_i $ and $\sum_{i=1}^{|\mathcal{V}|} \sum_{j=1}^{|\mathcal{A}|} X_{ij}\leq K$.
\begin{center}
\framebox[0.78\textwidth][c]{
\enspace
\begin{minipage}[t]{0.45\textwidth}
\small
\textbf{P.1:} \emph{Maximize $U(\mathcal{S})$}\\
\textbf{subject to:}
\begin{equation*}
\begin{cases}
\forall v_i \in \mathcal{V}:\mbox{    } \sum_{j=1}^{|\mathcal{A}|} X_{ij} \leq \kappa_i \mbox{ (\textbf{C1:} individual attention constraint)}\\
\sum_{i=1}^{|\mathcal{V}|} \sum_{j=1}^{|\mathcal{A}|} X_{ij}\leq K \mbox{ (\textbf{C2:} overall attention constraint)}\\
\forall v_i \in \mathcal{V}, a_j\in \mathcal{A}: X_{ij}\in \{0,1\}

\end{cases}
\end{equation*}
\end{minipage}
}
\end{center}
\begin{theorem}
\label{thm:np1}
The budgeted social advertising problem is NP-hard.
\end{theorem}
\emph{Proof:} We will reduce our problem from the traditional influence maximization problem. Let $\kappa_i=\infty$ for each $v_i$, then the only constraint is the overall attention constraint $K$. Assume there is only one ad, i.e., $|\mathcal{A}|$=1, our problem under this setting is equivalent traditional influence maximization problem \citep{kdd03}, i.e., finding a set of seeds that maximize the utility function $U(\mathcal{S})$ subject to cardinality constraint $K$. $\Box$

\subsection{Unconstrained Social Advertising Problem}
Given a seeds selection $\mathcal{S}$, we first define the cost function $C(\mathcal{S})$. \[C(\mathcal{S}) = \lambda_1 \cdot \underbrace{\sum_{i=1}^{|\mathcal{V}|}\exp(\max\{0, \sum_{j=1}^{|\mathcal{A}|} X_{ij}-\kappa_i\})}_{\mathrm{Part I}}+  \lambda_2 \cdot \underbrace{\exp(\max\{0, \sum_{i=1}^{|\mathcal{V}|} \sum_{j=1}^{|\mathcal{A}|}X_{ij} -  K\})}_{\mathrm{Part II}} \]
where Part I (resp. Part II) is the penalty resulting from exceeding the individual attention budget (resp. the overall attention budget),  $\lambda_1$ (resp. $\lambda_2$) is a parameter that determines how strictly we would like to penalize for exceeding the individual budget (resp. the overall budget). Then the objective function is defined as $U(\mathcal{S})-C(\mathcal{S})$.
\begin{center}
\framebox[0.78\textwidth][c]{
\enspace
\begin{minipage}[t]{0.45\textwidth}
\small
\textbf{P.2:} \emph{Maximize $U(\mathcal{S})-C(\mathcal{S})$}
%
\end{minipage}
}
\end{center}
\begin{theorem}
The unconstrained social advertising problem is NP-hard.
\end{theorem}
\emph{Proof:} Similar to the proof of Theorem \ref{thm:np1}, we still reduce our problem from traditional influence maximization problem. Consider the following problem setting: assume there is only one advertiser, and  $\lambda_1=0$, $\lambda_2=\infty$. Then one necessary condition to ensure the optimality for any given solution is to strictly obey the overall attention budget $K$. This problem is equivalent to the traditional influence maximization problem \citep{kdd03}, i.e., finding a set of seeds that maximize the utility function $U(\mathcal{S})$ subject to cardinality constraint $K$. $\Box$
\section{Budgeted Social Advertising Problem}
We instead turn to an alternative approach by first introducing a new utility function $V(\mathcal{S})$. Let \[V(\mathcal{S})=\sum_{a_i\in \mathcal{A}} V_i(\mathcal{S})\]
where \[V_i(\mathcal{S})= \min \{\alpha_i\cdot \sigma_i(S_i), B_i\}\]

Recall that $B_i$ is the maximum amount advertiser $a_i$ is willing to pay regardless of the seed selection, thus $V_i(\mathcal{S})$ can be interpreted as the actual payoff from $a_i$ under seed set $\mathcal{S}$. Now, we are ready to introduce the following \emph{revenue maximization problem} (RMP):

\begin{center}
\framebox[0.78\textwidth][c]{
\enspace
\begin{minipage}[t]{0.45\textwidth}
\small
\textbf{RMP:} \emph{Maximize $V(\mathcal{S})$}\\
\textbf{subject to:}
\begin{equation*}
\begin{cases}
\forall v_i \in \mathcal{V}:\mbox{    } \sum_{j=1}^{|\mathcal{A}|} X_{ij} \leq \kappa_i \mbox{ (\textbf{C1:} individual attention constraint)}\\
\sum_{i=1}^{|\mathcal{V}|} \sum_{j=1}^{|\mathcal{A}|} X_{ij}\leq K \mbox{ (\textbf{C2:} overall attention constraint)}\\

\end{cases}
\end{equation*}
\end{minipage}
}
\end{center}
\vspace{0.2in}

In order to find a solution for \textbf{P.1}, we first work with RMP and develop an algorithm with provable performance bound. Later, we slightly modify this solution, and obtain an approximation algorithm for \textbf{P.1}.

 Let $\widehat{\mathcal{S}^{\textbf{P.1}}} = \{\widehat{S^{\textbf{P.1}}_1}, \widehat{S^{\textbf{P.1}}_2}, \cdots, \widehat{S^{\textbf{P.1}}_{|\mathcal{A}|}}\}$ (resp. $\widehat{\mathcal{S}^{R}}=\{\widehat{S^{R}_1}, \widehat{S^{R}_2}, \cdots, \widehat{S^{R}_{|\mathcal{A}|}}\}$) denote the optimal solution of problem \textbf{P.1} (resp. RMP), i.e.,
\[\widehat{\mathcal{S}^{\textbf{P.1}}} =  {\arg \max}_{\mathcal{S}} U(\mathcal{S}), \mbox{                           }\widehat{\mathcal{S}^{R}}={\arg\max}_{\mathcal{S}} V(\mathcal{S})\] subject to  constraints $C1$ and $C2$. In the following lemma, we first establish a relation between \textbf{P.1} and RMP.
\begin{lemma}
\label{lem:11}
\[V(\widehat{\mathcal{S}^{R}}) \geq U(\widehat{\mathcal{S}^{\textbf{P.1}}})\]
\end{lemma}
\emph{Proof:} We prove this lemma through contradiction. First of all, since both $\mathcal{S}^*_{\textbf{P.1}}$ and $\mathcal{S}^*_{\mathrm{RMP}}$ must satisfy constraints $C1$ and $C2$, thus $\widehat{\mathcal{S}^{\textbf{P.1}}}$ and $\widehat{\mathcal{S}^{R}}$ are feasible solutions to both \textbf{P.1} and RMP. Assume by contradiction that $V(\widehat{\mathcal{S}^{R}}) < U(\widehat{\mathcal{S}^{\textbf{P.1}}})$, then for each $\widehat{S^{\textbf{P.1}}_i} \in \widehat{\mathcal{S}^{\textbf{P.1}}}$, either one of the following holds: (1) $\alpha_i\cdot\sigma_i(\widehat{S^{\textbf{P.1}}_i} )\leq B_i$, or (2) $\alpha_i\cdot\sigma_i(\widehat{S^{\textbf{P.1}}_i} )> B_i$. According to the definition of $U(\mathcal{S})$ and $V(\mathcal{S})$:
\begin{enumerate}
\item If (1) holds, $U_i(\widehat{\mathcal{S}^{\textbf{P.1}}}) = \alpha_i\cdot\sigma_i(\widehat{S^{\textbf{P.1}}_i} )$ and $V_i(\widehat{\mathcal{S}^{\textbf{P.1}}}) = \alpha_i\cdot\sigma_i(\widehat{S^{\textbf{P.1}}_i} )$;
\item Otherwise, if (2) holds, $U_i(\widehat{\mathcal{S}^{\textbf{P.1}}}) = 2\cdot B_i - \alpha_i\cdot \sigma_i(\widehat{S^{\textbf{P.1}}_i})$ and $V_i(\widehat{\mathcal{S}^{\textbf{P.1}}}) = B_i$.
\end{enumerate}
In either case, we have \[V_i(\widehat{\mathcal{S}^{\textbf{P.1}}})\geq U_i(\widehat{\mathcal{S}^{\textbf{P.1}}})\]
Then we have
\[V(\widehat{\mathcal{S}^{\textbf{P.1}}})=\sum_{i=1}^{|\mathcal{A}|} V_i(\widehat{\mathcal{S}^{\textbf{P.1}}}) \geq \sum_{i=1}^{|\mathcal{A}|} V_i(\widehat{\mathcal{S}^{\textbf{P.1}}}) \geq U(\widehat{\mathcal{S}^{\textbf{P.1}}})\]
Together with the assumption that $V(\widehat{\mathcal{S}^{R}}) < U(\widehat{\mathcal{S}^{\textbf{P.1}}})$, we have
\[V(\widehat{\mathcal{S}^{\textbf{P.1}}}) > V(\widehat{\mathcal{S}^{R}})\]
This contradicts to the assumption that $\widehat{\mathcal{S}^{R}}$ is an optimal solution of RMP. $\Box$

\begin{algorithm}[h]
{\small
\caption{Greedy-RMP (subroutine)}
\label{alg:greedy-peak1}
\textbf{Input:} Social network $\mathcal{G}$, budget $\mathcal{B}$, individual attention constraint $\kappa_i$, overall attention constraint $K$.\\
\textbf{Output:} Seed set $\mathcal{S}$.
\begin{algorithmic}[1]
\FOR {$S_i \in \mathcal{S}$}
\STATE $S_i = \emptyset$
\ENDFOR
\REPEAT
\STATE  without violating constraints $C1\sim C2$, add $v_i$ to $S_j$ that gives the highest marginal value in $V(\mathcal{S})$.
\UNTIL{no more seeds can be added without violating constraints $C1\sim C2$, or $V(\mathcal{S})$ has reached $\sum_{i=1}^{|\mathcal{A}|} B_i$.}
\RETURN $\mathcal{S}$.
\end{algorithmic}
}
\end{algorithm}

\begin{algorithm}[h]
{\small
\caption{Greedy-\textbf{P.1}}
\label{alg:greedy-peak}
\textbf{Input:} Social network $\mathcal{G}$, budget $\mathcal{B}$, individual attention constraint $\kappa_i$, overall attention constraint $K$.\\
\textbf{Output:} Seed set $\mathcal{S}$.
\begin{algorithmic}[1]
\STATE Call Algorithm \ref{alg:greedy-peak1} as a subroutine to find an initial seed set $\mathcal{S}$.
\FOR{$S_i \in \mathcal{S}$}
\IF {$V_i(S_i)=B_i$}
\STATE assume $u$ is the last seed that has been added to $S_i$ by Algorithm \ref{alg:greedy-peak1}
\IF {$U_i(S_i\backslash \{u\})\geq U_i(S_i)$}
\STATE $S_i = S_i\backslash \{u\}$
\ENDIF
\ENDIF
\ENDFOR
\RETURN $\mathcal{S}$.
\end{algorithmic}
}
\end{algorithm}
\begin{lemma}
\label{lem:111}
$V(\cdot)$ is a monotone submodular function.
\end{lemma}
\emph{Proof:} It is easy to prove that $V(\cdot)$ is a monotone function. We next prove the submodularity of $V(\cdot)$. The main idea is to first prove that $V_i(\cdot)$ is a submodular function, then the lemma follows from the fact that the sum of positive submodular functions is submodular.

Given two sets of seeds $X$ and
$Y$ such that $X\subseteq Y$, and consider the quantity
\[V_i(X \cup\{v\})- V_i(X)=\min \{\alpha_i\cdot \sigma_i(X\cup\{v\}), B_i\}- \min \{\alpha_i\cdot \sigma_i(X), B_i\}\]
\[V_i(Y \cup\{v\})- V_i(Y)=\min \{\alpha_i\cdot \sigma_i(Y\cup\{v\}), B_i\}- \min \{\alpha_i\cdot \sigma_i(Y), B_i\}\]
For ease of notation, let $\Delta_X = V_i(X \cup\{v\})- V_i(X)$ and $\Delta_Y = V_i(Y \cup\{v\})- V_i(Y)$.
\begin{itemize}
\item Case 1: $\alpha_i\cdot \sigma_i(X\cup\{v\})\geq B_i$ and $\alpha_i\cdot \sigma_i(X) \geq B_i$. This case is trivial $\Delta_X= \Delta_Y=0$.
\item Case 2: $\alpha_i\cdot \sigma_i(X\cup\{v\})\geq B_i$ and $\alpha_i\cdot \sigma_i(X) < B_i$. It follows that $\Delta_X= B_i - \alpha_i\cdot \sigma_i(X)$ and $\Delta_Y = B_i - \min\{\alpha_i\cdot \sigma_i(Y), B_i\} \leq B_i - \alpha_i\cdot \sigma_i(X) =  \Delta_X$.
\item Case 3: $\alpha_i\cdot \sigma_i(X\cup\{v\}) <  B_i$ and $\alpha_i\cdot \sigma_i(X) <  B_i$
\begin{itemize}
\item Case 3.1: $\alpha_i\cdot \sigma_i(Y\cup\{v\})< B_i$ and $\alpha_i\cdot \sigma_i(Y)< B_i$. We have $\Delta_X = \alpha_i\cdot \sigma_i(X\cup\{v\}) - \alpha_i\cdot \sigma_i(X)$ and $\Delta_Y = \alpha_i\cdot \sigma_i(Y\cup\{v\}) - \alpha_i\cdot \sigma_i(Y)$. Based on the fact that $\sigma_i(\cdot)$ is a submodular function \citep{kdd03}, we have $\Delta_X \geq \Delta_Y$.
\item Case 3.2: $\alpha_i\cdot \sigma_i(Y\cup\{v\})\geq B_i$ and $\alpha_i\cdot \sigma_i(Y)< B_i$. We have $\Delta_X = \alpha_i\cdot \sigma_i(X\cup\{v\}) - \alpha_i\cdot \sigma_i(X)$ and $\Delta_Y = B_i - \alpha_i\cdot \sigma_i(Y)\leq \alpha_i\cdot \sigma_i(Y\cup\{v\}) - \alpha_i\cdot \sigma_i(Y)$. Based on the fact that $\sigma_i(\cdot)$ is a submodular function, we have $\Delta_X \geq  \alpha_i\cdot \sigma_i(Y\cup\{v\}) - \alpha_i\cdot \sigma_i(Y) \geq \Delta_Y$.
\item Case 3.3:  $\alpha_i\cdot \sigma_i(Y\cup\{v\})\geq B_i$ and $\alpha_i\cdot \sigma_i(Y)\geq B_i$. We have $\Delta_X = \alpha_i\cdot \sigma_i(X\cup\{v\}) - \alpha_i\cdot \sigma_i(X)$ and $\Delta_Y=B_i-B_i=0$. Therefore, $\Delta_X \geq \Delta_Y$.
\end{itemize}
\end{itemize}
Therefore $\Delta_X \geq \Delta_Y$.  $\Box$
 \begin{lemma}
 \label{lem:matroid}
Given a finite ground set $\mathcal{X} = \{X_{ij}:0\leq i \leq |\mathcal{A}|; 0\leq j \leq |\mathcal{V}|\}$, and the independent sets $\mathcal{I}$ are defined as \[\mathcal{I} = \{T \subseteq \mathcal{X}: C1\mbox{ and }C2 \mbox{    are satisfied when }  X_{ij}=1 \mbox{ (resp. $X_{ij}=0$) for each } X_{ij}\in T \mbox{ (resp. $X_{ij}\notin T$)}\}\] then $(\mathcal{X}, \mathcal{I})$ is a matroid.
\end{lemma}
\emph{Proof:} It is easy to prove that property (I1) is satisfied, i.e., if $A \in \mathcal{I}$ and $B \subseteq A$, then $B\in I$.
We next prove that (I2) also holds. Let $\mathcal{X}_{*j}=\{X_{1j}, X_{2j}, \cdots, X_{|\mathcal{V}|j}\}$. If $A ,B \in \mathcal{I}$ and $|B|> |A|$, there must exist $j$ such that $|\mathcal{X}_{*j}\cap B| > |\mathcal{X}_{*j}\cap A|$, together with fact that $|A|<|B|\leq K$, and this means that adding any seed in $\mathcal{X}_{*j}\cap (B\backslash A)$ to $A$ will maintain independence, i.e., both C1 and C2 still hold. $\Box$
\begin{lemma}
\label{lem:22}
Algorithm \ref{alg:greedy-peak1} provides a 1/2-factor approximation for RMP.
\end{lemma}
\emph{Proof:} According to \citep{fisher1978analysis}, the greedy algorithm achieves 1/2-factor approximation for submodular maximization subject to one matroid constraint, then this lemma follows immediately from Lemma \ref{lem:111} and Lemma \ref{lem:matroid}. $\Box$
\begin{theorem}
\label{thm:main1}
Assume for each $a_i\in \mathcal{A}$, the minimum number of seeds needed to reach budget $B_i$ is at least 2, Algorithm \ref{alg:greedy-peak} provides a 1/4-factor approximation for \textbf{P.1}.
\end{theorem}
\emph{Proof:} Let $\mathcal{S}^{\mathrm{Alg}_1}=\{S^{\mathrm{Alg}_1}_1, S^{\mathrm{Alg}_1}_2, \cdots, S^{\mathrm{Alg}_1}_{|\mathcal{A}|}\}$ denote the seed set returned from Algorithm \ref{alg:greedy-peak1}, Lemma \ref{lem:22} indicates that $V(\mathcal{S}^{\mathrm{Alg}_1}) \geq V(\widehat{\mathcal{S}^{R}})$.  Notice that in Algorithm \ref{alg:greedy-peak}, we remove the last added seed $u$ from $\mathcal{S}^{\mathrm{Alg}_1}_i$ if and only if $V_i(S^{\mathrm{Alg}_1}_i)=B_i$, and $U_i(S^{\mathrm{Alg}_1}_i\backslash \{u\})\geq U_i(S^{\mathrm{Alg}_1}_i)$. We next prove that the total loss caused by removing all those seeds can be bounded.

Since the minimum number of seeds needed to reach budget $B_i$ is at least 2 for any ad $a_i$, we have $|S^{\mathrm{Alg}_1}_i|\geq 2$. Then based on the submodularity of  $V_i(\cdot)$ and the greedy manner of Algorithm \ref{alg:greedy-peak1}, we have \[V_i(S^{\mathrm{Alg}_1}_i\backslash \{u\}) \geq V_i(S^{\mathrm{Alg}_1}_i) - V_i(S^{\mathrm{Alg}_1}_i\backslash \{u\})\]
\[\Rightarrow V_i(S^{\mathrm{Alg}_1}_i\backslash \{u\}) \geq \frac{V_i(S^{\mathrm{Alg}_1}_i)}{2}\]
It follows that
\[V_i(S^{\mathrm{Alg}_1}_i\backslash \{u\}) = U_i(S^{\mathrm{Alg}_1}_i\backslash \{u\})\geq \frac{V_i(S^{\mathrm{Alg}_1}_i)}{2}\]
Let $\mathcal{S}^{\mathrm{Alg}_2}=\{S^{\mathrm{Alg}_2}_1, S^{\mathrm{Alg}_2}_2, \cdots, S^{\mathrm{Alg}_2}_{|\mathcal{A}|}\}$ denote the seed set returned from Algorithm \ref{alg:greedy-peak}, together with Lemma \ref{lem:22}, we have
\begin{eqnarray}
U(\mathcal{S}^{\mathrm{Alg}_2}) = \sum_{i=1}^{|\mathcal{A}|} U_i(S^{\mathrm{Alg}_2}_i) &=& \sum_{i\in H} U_i(S^{\mathrm{Alg}_1}_i\backslash \{u\}) + \sum_{i\in \mathcal{S}^{\mathrm{Alg}_2}\backslash H} U_i(S^{\mathrm{Alg}_1}_i)\\
&\geq& \sum_{i\in H} \frac{V_i(S^{\mathrm{Alg}_1}_i)}{2} + \sum_{i\in \mathcal{S}^{\mathrm{Alg}_2}\backslash H} V_i(S^{\mathrm{Alg}_1}_i)\\
&\geq& \frac{V(\mathcal{S}^{\mathrm{Alg}_2})}{2} \geq \frac{V(\widehat{\mathcal{S}^{R}})}{4}
\end{eqnarray}
 Based on Lemma \ref{lem:11}, we have \[U(\mathcal{S}^{\mathrm{Alg}_2}) \geq \frac{V(\widehat{\mathcal{S}^{R}})}{4} \geq \frac{U(\widehat{\mathcal{S}^{\textbf{P.1}}})}{4}\]
 This finishes the proof of this theorem. $\Box$

 \section{Unconstrained Social Advertising}
 In this section, we study the unconstrained social advertising problem. Notice that the original objective function  $U(\mathcal{S})-C(\mathcal{S})$  may take negative value, this may cause trouble for applying the concept of multiplicative approximation guarantee. To this end, instead of directly maximizing the original objective function, we equivalently maximize the following \emph{shifted} objective function \[f(\mathcal{S})=U(\mathcal{S})-C(\mathcal{S})+ \phi\] where $\phi$ is some constant to ensure $f(\mathcal{S}) \geq 0$ for any $\mathcal{S}$. In practise, we may choose $\phi$ as the maximum cost that can be incurred when allocating each ad to all users. In the rest of this section, we use $C^+(\mathcal{S})$ to denote $(C(\mathcal{S})+ \phi)$ for ease of notation.
\begin{algorithm}[h]
{\small
\caption{Randomized-U-RMP (subroutine)}
\label{alg:greedy-peak2}
\textbf{Input:} Social network $\mathcal{G}$, individual attention constraint $\kappa_i$, overall attention constraint $K$.\\
\textbf{Output:} Ad allocation $\mathcal{S}$.
\begin{algorithmic}[1]
\FOR {$S_t \in \mathcal{S}$}
\STATE $O_t = \emptyset$; $Q_t = \mathcal{V}$;
\FOR {$v_j \in \mathcal{V}$}
\STATE $a_i \leftarrow f'(O_t \cup \{v_j\})-f(O_t)$;
\STATE $b_i \leftarrow f'(Q_t \backslash \{v_j\})-f(Q_t)$;
\STATE $a'_i = \max\{0,a_i\}$; $b'_i=\max\{0,b_i\}$;
\STATE \textbf{with probability} $a_i'/(a_i'+b_i')$ \textbf{do}
\STATE $O_t \leftarrow O_t \cup \{v_j\}$
\STATE \textbf{else}
\STATE $Q_t \leftarrow Q_t \backslash \{v_j\}$
\ENDFOR
\ENDFOR
\STATE return $\mathcal{S}=\{O_1, O_2, \cdots, O_{|\mathcal{A}|}\}$
\end{algorithmic}
}
\end{algorithm}
\begin{algorithm}[h]
{\small
\caption{Greedy-\textbf{P.2}}
\label{alg:greedy-peak222}
\textbf{Input:} Social network $\mathcal{G}$, individual attention constraint $\kappa_i$, overall attention constraint $K$.\\
\textbf{Output:} Seed set $\mathcal{S}$.
\begin{algorithmic}[1]
\STATE call Algorithm \ref{alg:greedy-peak2} as a subroutine to find an initial seed set $\mathcal{S}$.
\FOR{$S_i \in \mathcal{S}$}
\IF {$V_i(S_i)=B_i$}
\STATE sort users in $S_i$ according to their marginal gains in terms of $V_i(\cdot)$ (same as in Algorithm \ref{alg:greedy-peak})
\STATE assume $u$ has the smallest marginal gain
\IF {$U_i(S_i\backslash \{u\})\geq U_i(S_i)$}
\STATE $S_i = S_i\backslash \{u\}$
\ENDIF
\ENDIF
\ENDFOR
\RETURN $\mathcal{S}$.
\end{algorithmic}
}
\end{algorithm}
 Similar to the approach used in the previous section, we first introduce the unconstrained revenue maximization problem (U-RMP) with the following objective function:
 \begin{center}
\framebox[0.78\textwidth][c]{
\enspace
\begin{minipage}[t]{0.45\textwidth}
\small
\textbf{U-RMP:} \emph{Maximize $f'(\mathcal{S}) = V(\mathcal{S})-C^+(\mathcal{S})$}
%
\end{minipage}
}
\end{center}
 Let $\widehat{\mathcal{S}^{\textbf{P.2}}}$ (resp. $\widehat{\mathcal{S}^{U}}$) denote the optimal solution of problem \textbf{P.2} (resp. U-RMP), i.e.,
\[\widehat{\mathcal{S}^{\textbf{P.2}}} =  {\arg \max}_{\mathcal{S}} f(\mathcal{S}), \mbox{                           }\widehat{\mathcal{S}^{R}}={\arg\max}_{\mathcal{S}} f'(\mathcal{S})\] Similar to Lemma \ref{lem:11}, we first prove that
\begin{lemma}
\label{lem:33}
\[f'(\widehat{\mathcal{S}^{R}}) \geq f(\widehat{\mathcal{S}^{\textbf{P.2}}})\]
\end{lemma}
The proof of Lemma \ref{lem:33} is similar to Lemma \ref{lem:11} thus omitted here.
 \begin{lemma}
 \label{lem:222}
 The function $f'(\mathcal{S})$ is submodular.
 \end{lemma}
 \emph{Proof:} To prove this lemma, it suffices to show that $V(\mathcal{S})$ is submodular and $C^+(\mathcal{S})$ is supermodular. The first part immediately follows from Lemma \ref{lem:111}.

 Now we focus on proving that $C^+(\mathcal{S})$ is supermodular. We expand $C^+(\mathcal{S})$ as follows \[C^+(\mathcal{S})  = \lambda_1 \cdot \underbrace{\sum_{i=1}^{|\mathcal{V}|}\exp(\max\{0, \sum_{j=1}^{|\mathcal{A}|} X_{ij}-\kappa_i\})}_{\mathrm{Part I}}+  \lambda_2 \cdot \underbrace{\exp(\max\{0, \sum_{i=1}^{|\mathcal{V}|} \sum_{j=1}^{|\mathcal{A}|}X_{ij} -  K\})}_{\mathrm{Part II}}+\phi\] It is easy to verify that both Part I and Part II are supermodular. Since nonnegative linear combination of supermodular functions is supermodular, then together with the fact that $\phi$ is a constant, we can prove that $C^+(\mathcal{S})$ is supermodular.

 This finishes the proof of this lemma. $\Box$

 Now we are ready to present a linear-time 1/2-approximation algorithm for U-RMP, which is adapted from \citep{buchbinder2012tight}. The detailed description can be found in Algorithm \ref{alg:greedy-peak2}. The algorithm maintains two candidate sets $O_t$ and $Q_t$ for each $S_t$. Initially, we set $O_t = \emptyset$; $Q_t = \mathcal{V}$. In each iteration we either adds $v_i$ to $O_t$ or removes it from $Q_t$. The decision is made randomly with probability derived from the marginal gain of each of the two options, i.e., $a'_i$ and $b'_i$. The algorithm terminates when $O_t$ and $Q_t$ are equal.
 \begin{lemma}
Algorithm \ref{alg:greedy-peak2} provides a 1/2-factor approximation for U-RMP.
\end{lemma}
\emph{Proof:}  According to Theorem I.2 in \citep{buchbinder2012tight}, the greedy algorithm achieves 1/2-factor approximation for unconstrained submodular maximization, then this lemma follows immediately from Lemma \ref{lem:222}. $\Box$
\begin{theorem}
Assume for each $a_i\in \mathcal{A}$, the minimum number of seeds needed to reach budget $B_i$ is at least 2, Algorithm \ref{alg:greedy-peak222} provides a 1/4-factor approximation for \textbf{P.2}.
\end{theorem}
\emph{Proof:}  Similar to  Theorem \ref{thm:main1}, we prove that the total loss can be bounded after removing some users from Algorithm \ref{alg:greedy-peak2}. Assume $u$ has been removed from $S_i$ in Algorithm \ref{alg:greedy-peak222}, based on the submodularity of  $V_i(\cdot)$ and the fact that $u$ has the smallest marginal gain, we have \[V_i(S^{\mathrm{Alg}_3}_i\backslash \{u\}) \geq V_i(S^{\mathrm{Alg}_3}_i) - V_i(S^{\mathrm{Alg}_3}_i\backslash \{u\})\]
\[\Rightarrow V_i(S^{\mathrm{Alg}3}_i\backslash \{u\}) \geq \frac{V_i(S^{\mathrm{Alg}_3}_i)}{2}\]
It follows that
\[V_i(S^{\mathrm{Alg}_3}_i\backslash \{u\}) = U_i(S^{\mathrm{Alg}_3}_i\backslash \{u\}) = U_i(S^{\mathrm{Alg}_4}_i) \geq \frac{V_i(S^{\mathrm{Alg}_3}_i)}{2}\]
On the other hand, removing any user can only decrease the cost $C^+(\mathcal{S})$. Then we have
\[f(\mathcal{S}^{\mathrm{Alg}_4})=U(\mathcal{S}^{\mathrm{Alg}_4}) + C^+(\mathcal{S}^{\mathrm{Alg}_4})\geq \frac{V(\mathcal{S}^{\mathrm{Alg}_3})}{2}+ C^+(\mathcal{S}^{\mathrm{Alg}_3})\geq \frac{f'(\mathcal{S}^{\mathrm{Alg}_3})}{2}\geq \frac{f'(\widehat{\mathcal{S}^{R}})}{4}\]
Then based on Lemma \ref{lem:33}, we have
\[f(\mathcal{S}^{\mathrm{Alg}_4})\geq \frac{f(\widehat{\mathcal{S}^{\textbf{P.2}}})}{4}\]
$\Box$
\section{Conclusion}
In this paper, we propose and study two social advertising problems: \emph{budgeted social advertising problem} and \emph{unconstrained social advertising problem}.  We first prove that both problems are NP-hard by reducing them from traditional influence maximization problem, then develop two constant factor approximation algorithms for each problem.

 \bibliographystyle{ormsv080}
\bibliography{social-advertising-1}
\end{document}